\input{aipcheck.tex}

\documentclass[
   ,final
  ]
  {aipproc}

\layoutstyle{6x9}
\usepackage{graphicx}

\begin{document}

\title 
      [The evolution of galaxies]
      {The Evolution of Galaxies:\\ an Infrared Perspective}


\author{Cristina C. Popescu}{
  address={Jeremiah Horrocks Institute, University of Central Lancashire,
  UK},
  email={cpopescu@uclan.ac.uk},
  thanks={}
}

\iftrue
\author{Richard J. Tuffs}{
  address={Max Planck Institut f\"ur Kernphysik, Germany},
  email={Richard.Tuffs@mpi-hd.mpg.de},
}

\fi

% \copyrightholder{Acoustical Society of America}
\copyrightyear  {2001}

\begin{abstract}
Understanding the infrared emission of galaxies is critical to observational 
and theoretical investigations of the condensation of galaxies out of the 
intergalactic medium and the conversion of gas into stars over cosmic time. 
From an observational perspective, about half of all photons emitted within 
galaxies are locally absorbed by dust grains, necessitating a self-consistent 
analysis of the panchromatic emission of galaxies to quantify star-formation
and AGN activity as a function of epoch and environment. From a theoretical 
perspective, physical processes involving dust are expected to play a major 
role in regulating the accumulation of baryons in galaxies and their 
condensation into stars on scales ranging from Mpc down to sub-pc. All this 
requires a quantitative analysis of the interaction between dust, gas and 
radiation. Here we review progress in the modelling of some of these processes.
 
\end{abstract}

\date{\today}

\maketitle

\section{Introduction}

Traditionally, the presence of dust in galaxies has been regarded as a 
nuisance preventing a clear view of stars in galaxies. This is even more true 
if one is interested in newly formed stars, where dust is not merely a 
nuisance but a show stopper. This is not only due to the enhanced absorption 
probability of UV photons radiated predominantly by the young stars, but also 
because of the strong spatial correlations between young stars and gas and 
dust in galaxies. Fortunately infrared space astronomy is allowing us in a 
real sense to hunt for the dark and uncover the obscured star formation, since 
the stellar photons that are obscured by dust become visible in the
infrared (IR). And in fact half of the energy emitted by all stars in the 
Universe since the Big Bang is absorbed by dust grains and is re-emitted in the
IR, as revealed by measurements of the extragalactic background. So  
understanding dust emission is crucial to the understanding of the 
star-formation history of the Universe, which is intimately connected to the 
understanding of the conversion of gas into stars in galaxies over cosmic time.

At the same time dust can 
influence structure formation in the Universe through even more direct
physical mechanisms affecting the thermodynamic state of gas inside and 
outside of galaxies, since dust is a primary coolant for the gas at all scales,
 in the intergalactic medium (IGM), in the interstellar medium (ISM) and in 
the star-forming clouds.

Here we shall review models for the conversion of stellar photons into
infrared emission as well as recent progress in our understanding of dust as a
main coolant for the gas. The first part is devoted to the mechanisms
through which dust can affect the thermodynamic state of the gas from Mpc to 
pc scales. The second part describes recent progress in our ability to 
quantitatively model measurements of the amplitude  and colours of dust 
emission to recover the nature of the stellar populations powering the 
infrared emission and other physical characterics of galaxies. 

Previous reviews given on this topic can be found
in \citep{ref:Popescu_Tuffs2009, ref:Popescu_Tuffs2008, ref:Popescu_Tuffs2007, 
ref:Popescu_Tuffs2005, ref:Dopita2005, ref:Madden2005, ref:Bianchi2004}.
 
\section{Dust regulating the thermodynamic state of the gas}
\label{sec:cooling}

\noindent
{\bf The intergalactic medium}

One of the biggest puzzles concerning galaxy evolution and baryonic physics on 
scales of hundreds of kpc is the mechanism by which galaxies can accrete gas to 
fuel the ongoing star-formation in their disks. One observes in the Milky Way 
halo and around other massive galaxies cold infalling HI clouds but these 
galaxies are supposed to reside in hot virialized haloes and the nature of the 
thermal instabilities needed to cool gas from temperatures of $10^6 - 10^7\,$K  
(for the more massive haloes) is unknown. However it is sometimes forgotten 
that the most efficient way of cooling such a medium is not through 
bremsstrahlung or line emission in the Xrays, but through inelastic collisions 
of the hot electrons and ions with dust grains, providing there are some small 
quantities of grains in this medium (see \citep{ref:Popescu_etal2000a}). 
Since collisional heating of dust grains 
is such an efficient mechanism, even trace quantities of grains can make a 
difference to the cooling for the gas, or at least trigger further cooling. 
These mechanisms have long been studied in the context of shocked interstellar 
gas (e.g. \citep{ref:DwekWerner1981}) but it is only recently that it was
realised that the effect of dust cooling should be considered in the context 
of structure formation.

This process has been recently investigated by \citep{ref:MontierGiard2004}, who
derived cooling curves for the gas in the IGM with and without the presence of
dust grains. Fig.~1 (left panel) shows their curves 
for different grain
sizes and for a dust-to-gas ratio of $10^{-4}$ . At low temperatures, between
$10^4 - 10^6\,$K, where the medium is not fully ionised, the cooling is 
dominated by atomic processes. For higher temperatures, $T>10^6\,$K, where up 
to now
bremsstrahlung was considered to be the most efficient cooling mechanism, it is
shown that the dust is the dominant cooling factor. 
\citep{ref:MontierGiard2004} also
calculated how the ratio between infrared to X-ray dominated cooling changes 
as a function of the dust-to gas ratio.  Fig.~1
(right panel) 
shows that for a dust-to gas ratio of $10^{-4}$ infrared cooling is more 
efficient for temperatures above $10^6\,$K, while for smaller dust-to gas ratios 
this condition applies at increasingly high temperatures. 

\begin{figure}
\caption{{\bf Left}: Cooling curves for gas in the IGM without dust (solid line) and
with dust (dotted and dashed curves for grain sizes of 0.5; 0.025 and
0.001\,${\mu}$m) from \citep{ref:MontierGiard2004}. The intergalactic gas is
modelled by a mixture of hydrogen and helium in cosmic proportion $X=0.75$ and
$Y=0.25$, respectively, with a dust-to-gas ratio of $Z_d=10^{-4}$ and without a
background UV radiation field. The cooling is due to recombination, collisional
ionisation, collisional excitation, bremsstrahlung and dust emission. 
{\bf Right:} Infrared to X-ray dominated cooling in the parameter space 
dust-to-gas ratio $Z_d$ and temperature of the gas, taken from 
\citep{ref:MontierGiard2004}.}
\includegraphics[height=.27\textheight]{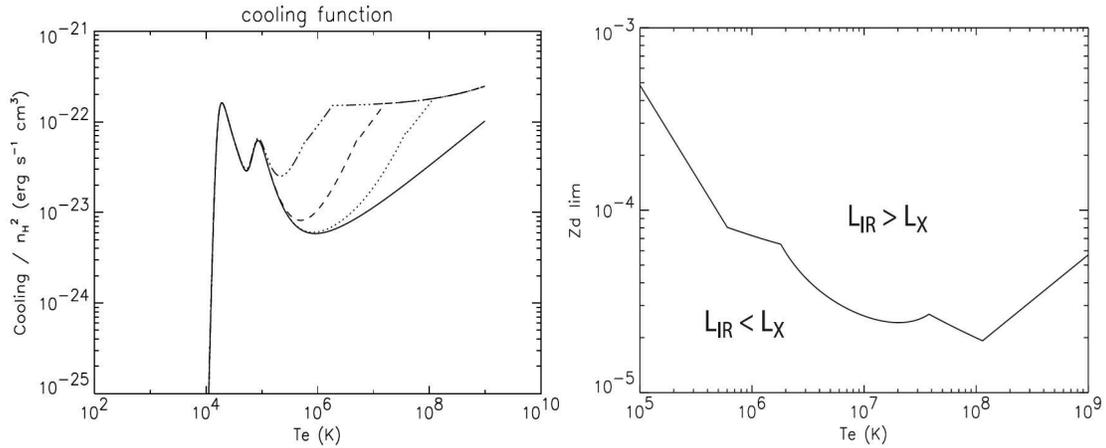}
\label{fig:montier04_cooling}
\end{figure}

The cooling functions described above have been coupled with numerical
simulations for galaxy formation by \citep{ref:Pointecouteau_etal2009}. Their
simulations showed that although the differences between the two cases (in the
absence and in the presence of dust) are not striking, the concentration of gas
in the cool and high-density phase is higher when dust is included than in the
purely gas cooling case.

To check whether these processes are operating in reality one should search
for dust emission associated with X-ray 
emission in the transition regions between the cosmic web filaments and the 
star-forming disks of galaxies and  ideal laboratories for this are clusters
and groups. Thus far, the best and perhaps only example  of a correspondence 
between FIR dust emission and X-ray emission from an IGM
around galaxies is in Stephan's Quintet (SQ) 
(see \citep{ref:Guillard_etal2010, ref:Natale_etal2010}), though it is unclear
  to what extent this correspondence is due to cooling instabilities
  in the IGM like those predicted in \citep{ref:MontierGiard2004}. Recent
  spectroscopic observations of the SQ (\citep{ref:Cluver_etal2010}) have 
  also shown  that the pure rotational lines of hydrogen can be effective in 
  cooling the IGM in the shock region of SQ. This was modelled as a 
  multi-phase medium by \citep{ref:Guillard_etal2009}.\\

\noindent
{\bf The interstellar medium}

Once inside galaxies, the subsequent evolution of the gas can
be affected by grains in other ways. As is well known, dust grains heat the 
diffuse ISM via the photoelectric effect and the thermodynamical balance is 
maintained through an equality between the photoelectric heating and the FIR 
cooling lines which are powered by inellastic collisions with gas particles.
Once the UV radiation field is suppressed, either by turning the sources off or 
by self-shielding by grains, the cooling is no longer balanced by heating, and 
the gas will condense further into denser structures, setting the seeds for 
condensation into dense molecular clouds.

In the simulation of \citep{ref:Juvela_etal2003} from Fig.~2 
one can see the predicted ratio between [CII] and FIR within an individual 
simulated cloud, where the darker regions delineate more optically thick 
regions of the cloud where the UV photons can't penetrate  but dust is still 
heated by the optical photons. Thus thermal pressure support can be 
reduced in density enhancement, potentially leading to further development of 
density contrast and ultimately affecting the propensity of gas to condense 
into stars. This shows again how dust is shaping the density structure of the 
gas, this time in the ISM.\\

\begin{figure}
\caption{Simulated image of the ratio between [CII] and FIR within an 
individual cloud, from \citep{ref:Juvela_etal2003}.}
\includegraphics[height=.25\textheight]{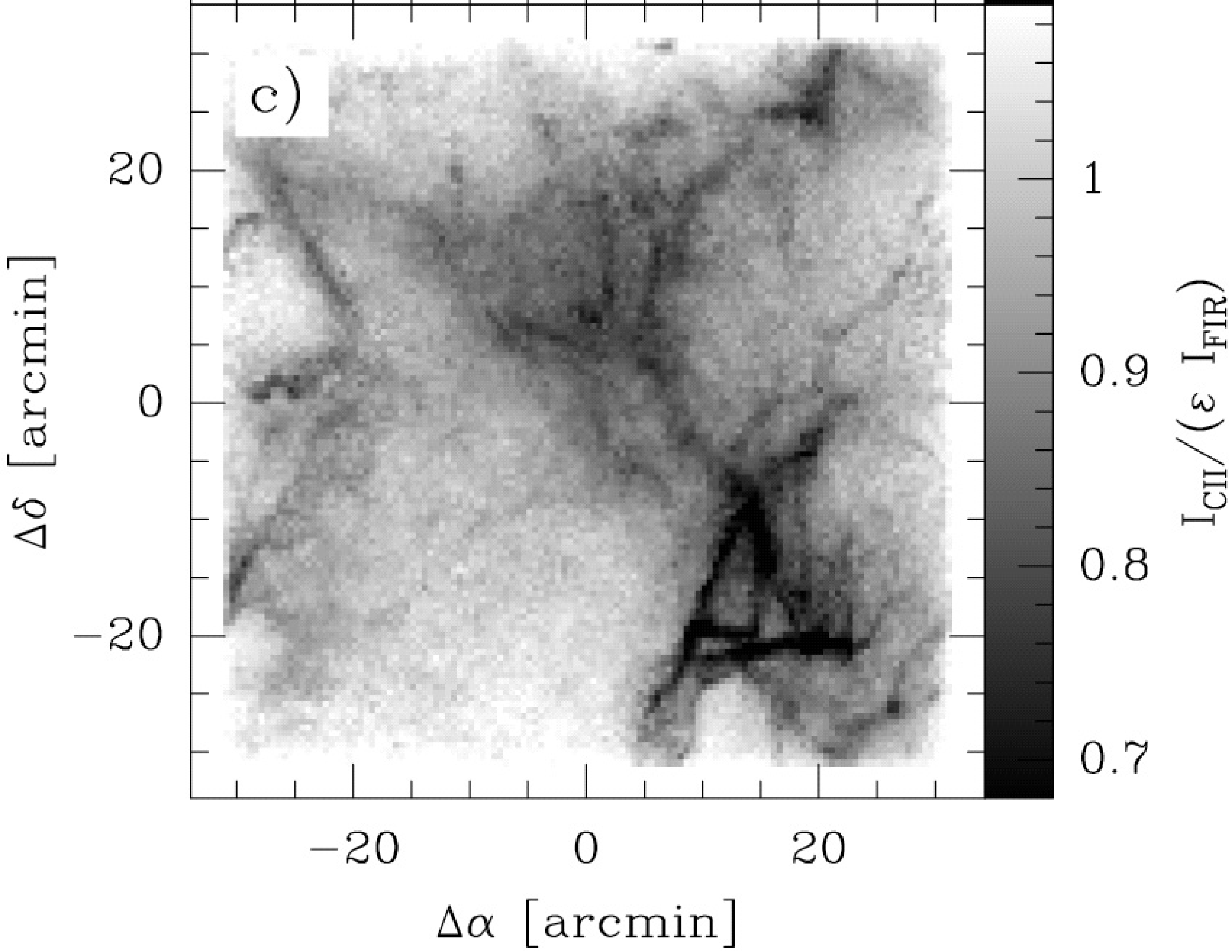}
\label{fig:juvela03}
\end{figure}

\noindent
{\bf The star-forming clouds}

It is well known that, except for the very first generation of stars, 
the cooling needed to precipitate the final stages of gravitational collapse in
star-forming regions is provided by inelastic collisions of molecules
with grains, visible through grain emission. 

Thus we have arrived at a logical end point of a journey showing
how dust physics might potentially influence the condensation of gas from 
the IGM into the ISM, then into denser structures within the ISM, and
finally into cloud cores and stars. Now we have to set the target:
- what are actually the star formation rates (SFRs) we have to ultimately
explain through all these processes? This is presented in the next section,
where we review recent techniques for modelling the conversion of stellar
photons into infrared photons in galaxies, which are the tools for modelling
the SEDs of galaxies to elucidate their physical parameters. 

\section{Modelling the SEDs of galaxies}

There are essentially two approaches to the problem of modelling the
transfer of radiation in galaxies and the interaction between dust particles
and photons. The first approach is to use multiwavelengths observations
as a starting point and build models that can be directly applied to the
panchromatic observations to decode their information. These are the SED model
tools, which perform a direct translation between observed quantities and 
physical quantities, for example to derive intrinsic distributions of stellar
 emissivity and dust in galaxies and to derive intrinsic physical parameters 
like dust opacity and SFRs. Then the
physical quantities can be compared with predicted physical quantities derived
from theory. We call this process {\it decoding observed panchromatic SEDs}:\\

\noindent
{\bf Observed SEDs --> Physical Quantities <--> 
Predicted Phys. Quantities <-- Theory}\\
{\bf \hspace*{2.9cm} | \hspace{3.9cm} |}\\
\hspace*{2.4cm} {\bf decode} \hspace{2.5cm} {\bf compare}\\ 

A second approach is to start with a theory, in other words to assume ``one 
knows  how galaxies form'', and make predictions for some physical quantities, 
e.g. to use as a starting point a simulation of a galaxy instead of an 
observation. Subsequently  a model that can deal with the dust physics and 
the transfer of radiation in the given simulation needs to be applied to obtain 
simulated  panchromatic images of  a galaxy for comparison with observations. 
The applied model is again a SED model tool, except that this works in the 
opposite direction from the previous approach. We call this process {\it encoding predicted physical
quantities}:\\

\noindent
{\bf Theory --> Predicted Physical Quantities --> 
Predicted SEDs <--> Observed SEDs}\\
{\bf \hspace*{7.3cm} | \hspace{3.4cm} |}\\
\hspace*{6.8cm} {\bf encode} \hspace{2.0cm} {\bf compare}\\ 

Obviously it is more difficult to decode than to encode. For example in the 
decoding process one needs to constrain the geometry of the problem, while in 
the encoding process the geometry is 
provided directly from theory. Nonetheless, the second approach relies on the 
assumption that the theory is correct, while in the first approach some
geometrical components can be constrained empirically. Since both approaches 
are needed, we would like both approaches to converge towards a unique 
solution.

\subsection{Decoding panchromatic SEDs}
\label{sec:sed}

There are several essential steps included in any self-consistent model for the
transfer of radiation and dust reprocessing. As mentioned before,
one of the essential steps is the specification of geometry, which means the 
specification of the distributions of stellar emissivity and dust, both on kpc 
scales and on pc scales. Components that are seen through optically thin lines
 of sight can in principle be empirically constrained from high
angular resolution optical observations. Those which are in very 
optically thick regime need to be fixed from physical considerations, with 
a posteriori consistency checks with data.
Optically thin components are the old stellar populations in the disk or in the
bulge. In edge-on galaxies \citep{ref:Xilouris_etal1998} has shown that in the 
optical and NIR bands one can derive the scalelength and scaleheights of the 
disk stellar
populations and associated dust, as well as the effective radius of the bulge
stellar populations by fitting observed images with simulated images produced
from radiative transfer calculations.

The geometry of young stellar populations and associated dust cannot be 
constrained directly from observations of direct stellar light, because young
stars are highly obscured in most cases. From physical
considerations we know that young stellar populations are born out of
molecular clouds and therefore initially have small velocity dispersions 
and small scaleheights. Furthermore we expect a strong spatial correlation
between the young stars and the dust associated with the parent 
molecular clouds. This situation has been  modeled
by \citep{ref:Popescu_etal2000b} by distributing the stellar
population in a thin exponential disk and the associated dust in a second 
layer which is also thin (small scaleheight) and occupies the same volume of 
space as the young stellar population. 
%The incorporation of the additional layer of dust with small scale-height has
%been succesfull in reproducing the detailed far-infrared
%spatial and spectral distribution of light in galaxies, as well as
%the attenuation-inclination relation derived for large statistical sample 
%(see \cite{ref:Driver_etal2007}.

Apart from specifying the geometry on kpc scales, an SED model needs to also
take into account the absorption of radiation on pc scales, which arises from
dust in the birth-clouds of massive stars. Because these clouds are spatially
correlated with their progeny on parsec scales, they are illuminated by a
strong UV-dominated radiation field of intensity 10-100 times that in the
diffuse ISM. This gives rise to a localised component of emission from grains
in thermal equilibrium with these intense radiation fields, which needs to be
properly modeled. In principle one would then need to design a
radiative transfer code that could operate from kpc down to pc scales. To date,
the best resolution achieved with an adaptive grid code when modelling the
observed SED of an individual galaxy is only 20\,pc (\citep{ref:Bianchi2008}), 
much coarser than the required pc resolution.   

For spiral galaxies we can circumvent this
problem by making the assumption that the heating of the grains in the birth 
clouds is dominated by photons from their stellar progeny, and by neglecting 
any external contribution from the diffuse ambient radiation fields in the 
modelled galaxy. This should be an excellent
approximation for spiral galaxies, where the filling factor of star-formation
regions is small. In this case we can perform the radiative transfer
calculations only for the diffuse component, and assume that a fraction $1-F$
of the radiation coming from the young stars in the birth clouds will escape
their clouds and propagate in the diffuse component, but essentially no photon 
from the diffuse component will be absorbed by and heat the dust inside the
clouds. This concept was introduced by the models of \citep{ref:Silva_etal1998} 
and \citep{ref:Popescu_etal2000b}. The low filling factor of opaque clouds in 
spiral disks assumed by this technique is supported by high resolution surveys of the Galactic Plane, e.g.
the APEX Telescope Large Area Survey (ATLAS) of the Galaxy at 870\,${\mu}$m
(\citep{ref:Schuller_etal2009}) . One should mention here that ATLAS is a very 
sensitive survey, going down to 50 mJy/beam, which is close to the $\tau=1$ 
limit in the V band, thus ensuring that any optically thick cloud would be 
detected in this survey. By contrast, the distribution of the CO line shows a
more diffuse distribution (see \citep{ref:Matsunaga_etal2001}), with large clouds 
overlapping and a very high  filling factor. This is because the 
$^{12}{\rm C}^{16}{\rm O}$ line 
traces the optically thin outer surfaces of the molecular clouds, which is a 
different volume of space than that traced by the 870\,${\mu}$m emission. We 
therefore caution modellers of the SEDs of spiral galaxies
not to follow the distribution of the CO emission when tracing the
distribution of dust clumps which are optically thick in the UV/optical.

For starburst galaxies the approximation of a small filling factor for the
star-formation regions is obviously no longer valid, and other procedures need
to be invoked when modelling this type of galaxies. One way is to consider 
these galaxies as a collection of HII regions, as modelled by 
\citep{ref:Dopita_etal2005} with the updates from \citep{ref:Groves_etal2008}. 
These models have a physical description for the evolution of the star-forming 
complexes and they also take into account collective effects due to merging of 
HII regions. However they still do not consider heating of the clumps by the 
external ambient radiation fields and do not have a diffuse 
component to contribute to the FIR emission. Another model proposed for 
starbursts is the hot spot model of \citep{ref:Siebenmorgen_Krugel2007}. This 
model considers star-forming clouds embedded in a diffuse radiation field, all 
within spherical symmetry, which is a reasonable assumption for starbursts. 
They also consider the heating of the clumps by the external radiation field, 
but this is not done in a self-consistent way. 

\begin{figure}
\caption{Left: Example of radial profiles of radiation fields 
(calculated using the model of \citep{ref:Popescu_etal2000b}) in the 
plane of a typical spiral galaxy having a bulge-to-disk ratio
$B/D=0.33$. Right: Predicted SEDs for different pressure of the ambient 
interstellar medium, from \citep{ref:Dopita_etal2005}.}
\includegraphics[height=.20\textheight]{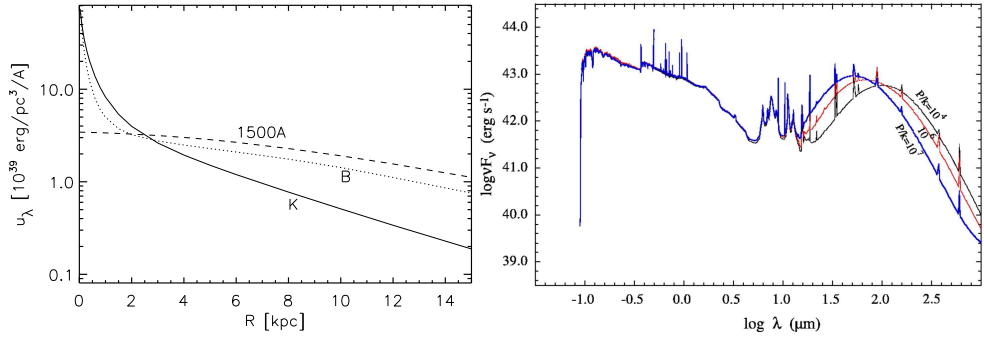}
\end{figure}

Having specified the geometry of the problem the next step in any SED model is
to run radiative transfer calculations (see \citep{ref:Kylafis_Xilouris2005} 
for a comprehensive review on radiative transfer techniques) to derive the 
radiation fields in
galaxies. Fig.~3 (left panel) shows an example of calculated radial 
profiles in the plane of a typical spiral galaxy having a bulge-to-disk ratio 
$B/D=0.33$. We note here the large variation in the colour of the
radiation fields with position, which will introduce large differences in the 
shape of the FIR SEDs, as well as in the predictions about the
role different stellar populations have in heating the dust as a function of
position in the galaxy. Finally this will strongly affect the conclusions
about the SFR and SF efficiencies in galaxies. Thus, models that assume 
radiation fields with the fixed colour  of the  local interstellar radiation 
fields (LIRF) are likely to 
introduce systematic uncertainties in the predictions for the dust emission 
SEDs and therefore for the conversion of gas into stars as a function
of galactocentric radius.

\begin{figure}
\caption{Temperature distributions for dust grains of different sizes (plotted
  as different curves in each panel) and various composition: Si (left
  panels), Gra (middle panels) and PAH$^{+}$ (right panels), heated by the
  diffuse radiation fields. The calculations were done for a typical spiral 
  galaxy having a bulge-to-disk ratio $B/D=0.33$ and using the model
  of \citep{ref:Popescu_etal2000b}. Temperature distributions for PAH$^{0}$ are 
  not plotted in this figure. Dashed lines are for grains with radius 
  $a< 0.001\,{\mu}$m (0.00035, 0.00040, 0.00050, 0.00063 and 
  0.00100\,${\mu}$m), dotted lines are for grains with $0.001<a\le0.01$ 
  (0.00158, 0.00251, 0.00398, 0.00631, 0.01000${\mu}$m) and solid lines are 
  for grains with $a>0.01$ (0.0316, 0.10000, 0.31623, 0.7943\,${\mu}$m). The 
  biggest grains have delta function distributions, since they emit at 
  equilibrium temperature. Going from the top to the bottom panels the 
  calculations are done for different positions in the model galaxy: 
  $r=0$\,pc, $z=0$\,pc; $r=15000$\,pc, $z=0$\,pc; $r=0$\,pc, $z=600$\,pc; and 
  $r=15000$\,pc, $z=600$\,pc. }
\includegraphics[height=.62\textheight]{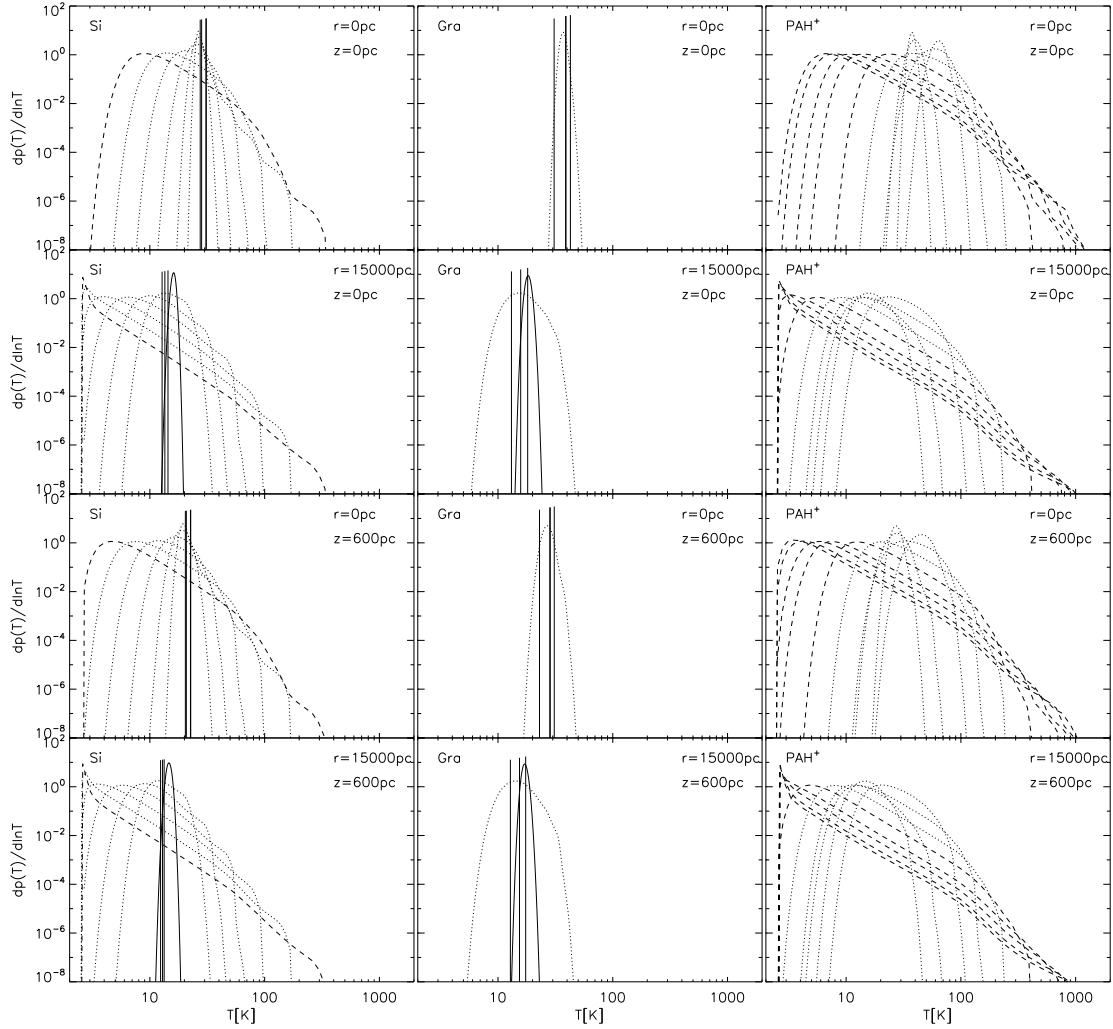}
\label{fig:temp}
\end{figure}

Once radiation fields in galaxies are calculated one can derive the temperature
distribution of grains of different sizes and composition as a function of
position in the galaxy. Here it is important to recall that most of the
grains in the interstellar medium are not heated in equilibrium with the
radiation fields, but are heated stochastically. Fig.~4 shows  
examples of calculated probability distributions of temperature for grains of 
different sizes, composition and at different positions in a galaxy. The same 
grain sizes are
plotted for each position and the calculation corresponds to a typical spiral
galaxy having a $B/D=0.33$. Overall small grains are more
stochastically heated and therefore exhibit large temperature fluctuations 
while big grains emit at equilibrium temperature and therefore exhibit
distributions close to or delta functions.  Apart from the dependence on grain
size, the temperature distributions also strongly depend on 
the intensity and colour of the radiation fields. One can see from
Fig.~4 that big grains placed in the centre of the galaxy, where 
radiation fields are stronger and redder will emit close to equilibrium 
temperature, while the same grains placed in the weaker and bluer radiation 
fields in the outskirts of the galaxy will exhibit temperature fluctuations. 
This emphasises the need to have a self-consistent calculation, where the color
and intensity of the radiation fields are calculated as a function of position
in the galaxy.

\begin{figure}
\caption{Spatially integrated dust and PAH emission SEDs of the diffuse
component calculated using the model of \citep{ref:Popescu_etal2000b}. 
Each panel shows the effect of changing one parameter of the model at a time, 
while the other parameters are kept fixed to the value corresponding to the 
best fit of NGC~891.}
\includegraphics[height=.48\textheight]{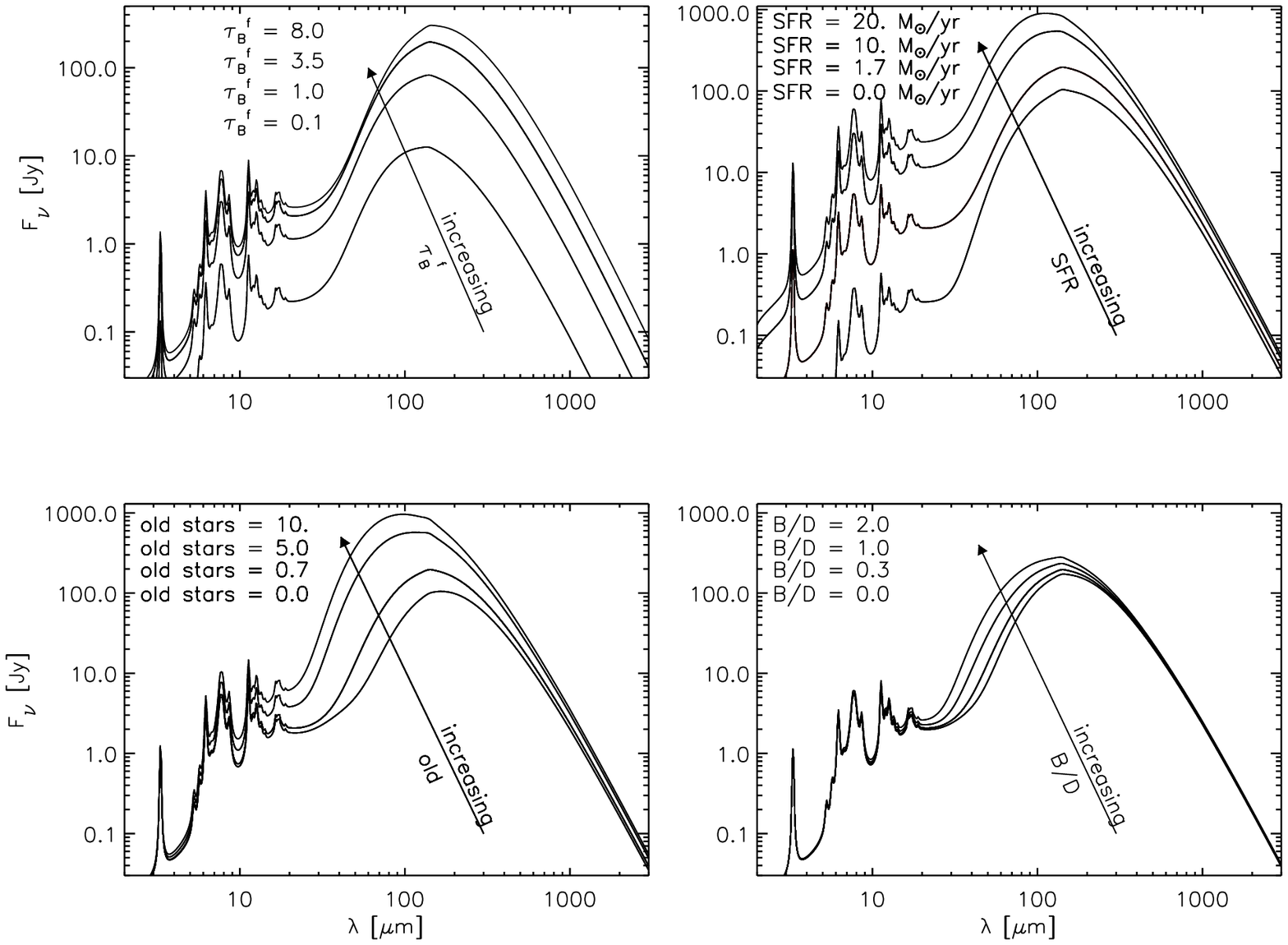}
\label{fig:sed}
\end{figure}

The spatially integrated dust and PAH SEDs can then be obtained by integrating
over all positions in a galaxy. We first show the predictions for the diffuse
component in spiral galaxies, using the model of \citep{ref:Popescu_etal2000b},
where the parameters of interest are: dust opacity ($\tau_B^f$), luminosity of 
the young stellar populations ($SFR$), luminosity of the old disk stellar 
populations ($old$)  and $B/D$. Fig.~5 shows SEDs where one 
parameter at a time is varied, all the others being fixed to the values
corresponding to the best fit of NGC~891. The main effect of increasing 
opacity is to increase the overall amplitude of the dust and PAH emission 
SEDs, with a fast increase for the optically thin cases and a slower increase 
and eventually a saturation for the optically thick cases.  We note here that 
the opacity is the only parameter that changes the submm level. Essentially 
one cannot increase the level of submm emission by just increasing the 
luminosity  of the heating sources. The effect of increasing the $SFR$ is to
produce warmer SEDs, both because the peak of the SEDs shifts towards shorter
wavelengths, but also because the ratio between mid-infrared (MIR) and FIR 
increases. 
An increase in the contribution of the old disk stellar
population $old$ produces a similar shift of the peak of the FIR SEDs towards 
shorter wavelengths, but the amplitude of the MIR/PAH emission does not increase
accordingly, as the PAH are mainly excited by the UV photons predominantly
produced by the young stellar population. The variation in the
$B/D$ produces very similar changes in the IR SEDs to those produced by the
variation in the disk old stellar
populations, but with a small dynamic range. Overall the ratio
between the PAH and FIR emission will be highly dependent on the ratio of the 
old to young stellar populations.

From the trends in Fig.~5 one can see
that overall the 4 parameters of the model are fairly orthogonal and it is 
therefore possible  to decode the FIR SEDs in a fairly non-degenerate way. Of
course one would need to add the clumpy component to the diffuse one to
obtain the total infrared SEDs, and this would add another parameter ($F$
- the clumpiness factor in the model of \citep{ref:Popescu_etal2000b}). 
However, the optical and UV data need also to be added 
when trying to decode the panchromatic SEDs. Thus, the corresponding 
predictions for attenuation of stellar light (\citep{ref:Tuffs_etal2004}, but
see also \citep{Pierini_etal2006} for alternative models) 
should be used in conjunction with the dust emission SEDs in order to obtain 
final solutions. The derived values of opacity can then be used to correct
surface-brightness (SB) photometry of stellar light for the effect of dust 
using corresponding models (see \citep{ref:Mollenhoff_etal2006}). Alternative
methods for obtaining dust opacities rely on statistical determinations (e.g. 
\citep{ref:Driver_etal2007}); see \citep{ref:Graham_Worley2008} for studies of
dust corrected SB photometric parameters.

Lets now consider the total IR emission from starburst galaxies. 
In the model of \citep{ref:Dopita_etal2005} (see also 
\citep{ref:Groves_etal2008}) the
infrared emission is taken to be exclusively emitted
by dust in PDR regions at the interface of HII regions
with their parent molecular clouds. The main factor
controlling the shape of the dust emission SED is then
the radius of the PDRs, which is larger for lower
density, low pressure environments, and smaller for
high density, high pressure environments. In 
Fig.~3 (right panel) one sees the effect of changing ambient pressure
from $P/k=10^4\,{\rm cm}^{-3}\,{\rm K}$ (ambient medium of a spiral galaxy) to 
$10^6$ through $10^7\,{\rm cm}^{-3}\,{\rm K}$ (ambient medium of a starburst).
The PDR is closer to the exciting stars for the higher pressure environment
and thus the dust emission is warmer for this case.

In the model of \citep{ref:Siebenmorgen_Krugel2007}, the IR
emission is emitted by a combination of clouds with
embedded stars and a diffuse dusty intercloud medium.
Thus, the temperature of the dust emission depends not
only on the compactness of the clouds, but also on the
opacity of the diffuse intercloud medium and the overall
luminosity density of the stars. \citep{ref:Siebenmorgen_Krugel2007} showed 
the effect of increasing the illumination of dust in the diffuse medium
by either increasing the number of stars or
by decreasing the radius of the starburst region
or by decreasing the opacity of the diffuse medium. By contrast, 
the effect of changing the size of the clumps (done in this model by a density 
parameter), was found to be quite small.

In conclusion, different models invoke different physical
mechanisms to change the illumination of grains in starburst
galaxies. The main difference relates to the relative
importance of the clumps and diffuse interclump
medium, and better observational constraints are needed to
pin down this issue. 

\subsection{Encoding predicted physical quantities}

As mentioned before, the second approach to modelling the SEDs of galaxies is 
that of encoding predicted physical quantities. In most cases this entails 
using simulated images of galaxies as input for dusty
radiative transfer codes to predict the appearance of these images as
would be seen through dust and also to predict the dust emission 
images - see \citep{ref:Jonsson_etal2009, ref:Chakrabarti_etal2008}. 
For example
\citep{ref:Jonsson_etal2009} uses the same approximation for the treatment of 
the clumpy component as that used in the models that decode the observed SEDs 
of spiral galaxies (see \cite{ref:Silva_etal1998} and 
\cite{ref:Popescu_etal2000b}). Thus, \citep{ref:Jonsson_etal2009}
 makes the assumption that the illumination of the
birth clouds is dominated by emission coming from the embedded stars, and
considers the radiative transfer only for the diffuse component, while the
clumpy component is treated separately, using the model of 
\citep{ref:Groves_etal2008}. So in this way the local absorption on pc scales 
is dealt with.

In this approach the geometry does not need to be specified, since it
is provided by the simulations. The use of simulations has the advantage to
provide quite spectacular images of galaxies (see 
Fig.~6). However, because of the high 
resolution required to produce all the structure provided by the simulations, 
explicit calculation of the radiation fields is 
not performed, as this would add an extra dimension to the overall 
calculation, and make them intractable. Since radiation
fields are not calculated, self-consistent calculations of the 
stochastic heating of the small grains and PAH molecules cannot be performed. 
In the model of \citep{ref:Jonsson_etal2009} the phenomenological model of 
\cite{ref:Draine_Li2007} is used instead to re-radiate the emission from 
stochastically heated grains.\\ 

\begin{figure}
\caption{Simulated images of a dust attenuated galaxy seen at different 
inclinations, from \citep{ref:Jonsson_etal2009}.}
\includegraphics[height=.15\textheight]{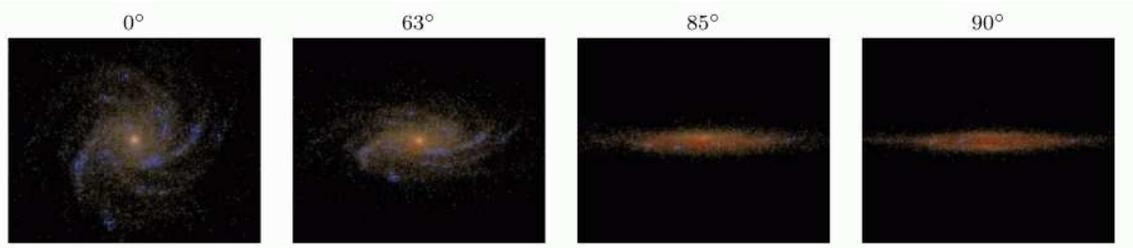}
\label{fig:jonssonetal2009}
\end{figure}

\noindent
Here we have attempted to give a broad classification of the different
approaches to the SED modelling of galaxies. It was our intention to introduce
general concepts rather than cover detailed models in the literature. A more
complete list of SED models also includes:  
\citep{ref:Siebenmorgen_Krugel1992, ref:Rowan-Robinson_Efstathiou1993, 
ref:Krugel_Siebenmorgen1994, ref:Efstathiou_etal2000, 
ref:Bianchi_etal2000, ref:Efstathiou_Rowan-Robinson2003, 
ref:Galliano_etal2003, ref:Takagi_etal2003,
ref:Misiriotis_etal2004, ref:Baes_etal2005, 
ref:Takeuchi_etal2005, ref:Misselt_etal2001,
ref:Piovan_etal2006}, where here we did not include models for AGN
tori.

\section{Outlook}

One of the biggest challenge for the SED models is to make them applicable to
the large statistical samples of panchromatic data that are now becoming
available for the local universe galaxies, like for example the GAMA survey
(\citep{ref:Driver_etal2009}), the Herschel ATLAS survey
(\citep{ref:Eales_etal2009}) and GALEX (\citep{ref:Martin_etal2010}). 
In the future we will also need to re-calibrate
our SED models to make them applicable to the distant universe, using data from
the next generation observatories, like SPICA (\citep{ref:Swinyard_etal2009}), 
ALMA, JWST and VLT. At the same time we will need to include 
dust physics in the simulations for galaxy formations within a fully
cosmological context. Realistic simulations of disk galaxies are now being for
the first time produced (see \citep{ref:Brook_etal2010}) and, within a few
years, we expect this to be routinely done. It is therefore crucial to
incorporate all known physics related to sources and sinks of grains in
galaxies and in the surrounding IGM, as well as the cooling and heating
mechanisms involving grains. Only then we will be able to make reliable 
predictions, which will directly constrain theories for the formation and 
evolution of galaxies.

% choose bibtex style depending on layout style and options used in
% sample:

%\doingARLO[\bibliographystyle{aipproc}]
%          {\ifthenelse{\equal{\AIPcitestyleselect}{num}}
%             {\bibliographystyle{arlonum}}
%             {\bibliographystyle{arlobib}}
%          }

\end{document}